\documentclass[12pt]{article}
\usepackage{a4}
\usepackage{graphicx,psfrag,rotating,subfigure}
\usepackage{float}
\usepackage{citesort}
\begin{document}
\vspace{15cm}

\title{
\vspace{2cm}
  {\huge
    Cherenkov radiations from Quark Plasma
  }
}

\vspace{10cm}

\author{ Kensuke Homma
\\
Physical Science,
Graduate School of Science,
Hiroshima Univ.
\\
1-3-1 Kagamiyama, Higashi-hiroshima 739-8526, Japan
\\
}

\date{September 11, 2003}
\maketitle

\begin{abstract}
I propose an observable to declare the formation of
Quark Plasma (QP) without any doubts.
It is the observation of Cherenkov photon radiations
of the order of 100 MeV associated with
fast light charged particles while they are penetrating
hot dense mediums produced in high energy heavy ion collisions.
Direct observations of Cherenkov rings associated
with charged leptons decayed from
$Z^0$ boson or Drell-Yan process which must exist
earlier than the QCD medium formation
and photon emissions associated with high $p_t$ hadrons
would be a definite signature of QP formation.
\vspace{1.0cm}
keywords    : quark, gluon, plasma, Cherenkov, photon, radiation
\end{abstract}

\newpage
Many observables have been proposed
to declare the formation of Quark-Gluon Plasma (QGP) such as
strong elliptic flow\cite{STAREF130}\cite{PHENIXEF200}, jet quenching
\cite{PHENIXJetSup130}\cite{STARJetSup130}\cite{PHENIXJetSupPi0200}\cite{PHENIXJetSupCh200}, 
$J/\psi$ suppression\cite[references are therein]{NA50:2003} and
increase of thermal photon yields or thermal dilepton yields
\cite[references are therein]{WA98:2000}.
However, none of above observables can make a definite
statement by itself without assumptions or
parameter tunings based on preferable models. 
One of reasons which makes it difficult to conclude is
the complexity of strong interactions at the initial and final states
involved in the observables.
Although thermal photons, in principle, infer partially the electromagnetic
interactions in the plasma, it is experimentally non trivial
to measure the slight increase of the total photon yield compared to
huge background from neutral pion decays\cite{WA98:2000}. In addition, it can not
conclude from the experiment itself that the radiations are definitely
not coming from the hadron phase after all\cite{DirectPhotonTwoLoop}.
In this letter, apart from the strong interactions,
I would like to propose another observable
to declare that Quark Plasma has been evidently formed by not allowing
any other interpretations.

Since quarks can carry electrical charges as well as color charges,
the analogy to measurements of a refractive index
in a classical electromagnetic medium is worth reconsidering.
Let us remind of the plasma frequency $\omega_p$ based upon classical
electromagnetism (any text books can be referred, see \cite{JACKSON} for instance):
\begin{equation}\label{eqn:wp2}
\omega_p^2 = 4\pi \alpha {e_q}^2 \frac{N_q}{m_q}
\end{equation}
where $N_q$ is the number density of
quarks, $m_q$ is a mass of quark, $e_q$ is the electrical charge
in unit of electron charge and $\alpha=1/137$.
If asymptotically free quarks are assumed in an ideal gas and
we focus on the aspect of electromagnetic plasma in QGP,
the relation above is still valid.
Suppose that genuine confinement or restoration of chiral symmetry occurs
in the heavy ion collisions, the electromagnetic plasma frequency of the dense medium
must become higher, because the masses of quarks are almost zero as expected
from Eq.(\ref{eqn:wp2}).
Although this might be a too simplified case, it is still instructive
to emphasize the direct relation between the plasma frequency and
the characteristic mass of carriers in the plasma.
If we assume u-quarks as dominant carriers of the plasma
with its temperature $T \simeq 150 - 200$ MeV,
one can estimate the number density based upon the Fermi-Dirac distribution
in a relativistic ideal gas as
\begin{equation}\label{eqn:Nu}
N_u = N_{\bar{u}} = \frac{g_u}{2\pi^2} 
                    \int_{0}^{\infty} \frac{p^2 dp}{1+exp(\sqrt{p^2+m_q^2}/T)}
                  \simeq 0.73 - 1.72 [fm^{-3}]
\end{equation}
where $p$ is the momentum,
$g_u= 2 \times 9$ is the degree of freedom of u-quark and
$m_q$ is the current u-quark mass of 5 MeV.
On the other hand, if we assume pions as dominant carriers of the plasma
with the same temperature range,
we can estimate the number density based upon the Bose-Einstein distribution
in a relativistic ideal gas as
\begin{equation}\label{eqn:Npi}
N_{\pi^+} = N_{\pi^-} = \frac{g_{\pi}}{2\pi^2} 
                        \int_{0}^{\infty} \frac{p^2 dp}{exp(\sqrt{p^2+m_{\pi}^2}/T)-1}
                      \simeq 0.04 - 0.11[fm^{-3}]
\end{equation}
where $g_{\pi}$ is 1 and $m_{\pi}$ is the mass of charged pion.
With Eq.(\ref{eqn:wp2}), (\ref{eqn:Nu}) and (\ref{eqn:Npi})
we can estimate the electromagnetic plasma frequency $\omega_p(u\bar{u})$ and
$\omega_p(\pi^{\pm})$ in the $u\bar{u}$ and $\pi^+\pi^-$ plasma as;
\begin{equation}\label{eqn:wp}
\omega_p(u\bar{u})
         = \sqrt{ 4\pi \alpha (e_u^2 + e_{\bar{u}}^2) \frac{N_u}{m_u} }
         \simeq 238 - 366 [MeV]
\end{equation}
and 
\begin{equation}\label{eqn:wppi}
\omega_p(\pi^{\pm})
         = \sqrt{ 4\pi \alpha (e_{\pi^+}^2 + e_{\pi^-}^2) \frac{N_{\pi}}{m_{\pi}} }
         \simeq 20 - 33 [MeV].
\end{equation}
What should be stressed here is that there is a clear distinction between
Eq.(\ref{eqn:wp}) and (\ref{eqn:wppi}) due to $m^{-1/2}$ dependence.
This is an essentially important fact for the following discussions.

In order for Cherenkov radiation to be
observed, following necessary conditions must be satisfied;
\begin{equation}\label{eqn:cherenkov}
\delta n(\omega) = n(\omega) - 1 > 0
\end{equation}
\begin{equation}\label{eqn:cos}
cos\theta = 1/\beta n(\omega)
\end{equation}
where $\omega$ is a frequency to discuss the dispersion property,
$n(\omega)$ is a refractive index of a dense medium,
$\delta n(\omega)$ is the deviation of it from bear vacuum,
and $\theta$ is the emission angle from the direction of a fast charged particle.
In order to discuss frequencies of emitted Cherenkov photons, let us remind
that a dielectric constant $\epsilon(\omega)$ of an electromagnetic medium has a simple form:
\begin{equation}
n(\omega)^2 = \epsilon(\omega) = 1 + \frac{\omega_p^2}{\omega_R^2 - \omega^2}
\end{equation}
where $\omega_R$ is a resonant frequency.
It is natural to expect that some resonances appear in loosely
bound quarks or hadrons during the phase transition from a deconfined to confined state.
Here the dumping term in the more general form was omitted for the simplicity,
by which the intention of this letter is not altered.
In a high frequency limit $\omega >> \omega_R$, the relation becomes
\begin{equation}
n(\omega)^2 = \epsilon(\omega) = 1 - \frac{\omega_p^2}{\omega^2}
\end{equation}
where Cherenkov emissions can not be expected due to $\delta n(\omega)<0$.
As the frequency approaches $\omega \simeq \omega_R$, the anomalous dispersion appears
and $\epsilon(\omega)$ steeply fluctuates around unity.
Below the resonant frequency $\omega < \omega_R$, $\delta n(\omega)>0$
is satisfied and Cherenkov radiations may occur.
It is a well-know fact that the frequencies of emitted Cherenkov photons
$\omega_C$ are limited in a frequency band $\omega_0 \leq \omega_C < \omega_R$,
where $\omega_0$ must satisfy
$n(\omega_0) > 1/\beta$\cite{JACKSON}. Since the plasma frequency is higher than the
resonance frequency, if some of observed Cherenkov photons satisfy the
following relation
\begin{equation}
\omega_p(\pi^+\pi^-) << \omega_c < \omega_R < \omega_p(u\bar{u}) ,
\end{equation}
it directly indicates that the dominant carriers of the plasma are quarks,
because the hadronic matter can not emit Cherenkov photons exceeding $\omega_p(\pi^{\pm})$.
This is a definite signature to declare the formation of Quark Plasma as well as
the restoration of the chiral symmetry. In other words, this measurement provides us
a direct way to measure the bare quark mass near the phase transition.
In addition, if we could measure the emission angles with respect
to fast light charged particles as well as those energies, we can directly measure the
refractive index by experiments and eventually investigate
the dispersion property of the hot dense medium.
With $\beta = \sqrt{1-(\frac{m_{ch}}{E_{ch}})^2}$
where $m_{ch}$ and $E_{ch}$ are mass and total energy of a fast charged particle,
the sensitivity to $\delta n$ can be discussed by the relation
\begin{equation}
\delta n > 1/\beta - 1.
\end{equation}
As the penetrating light charged particles,
if we take electrons from $Z^0 \rightarrow e^+ e^-$ which must exist
much earlier than the QCD medium formation,
it is sensitive to $\delta n > 6.0 \times 10^{-11}$ in a produced medium.
If we take hard struck u-quarks of 10 GeV/c and $m_q=5$MeV
as sources of high $p_t$ hadrons, it is sensitive to
$\delta n > 1.3 \times 10^{-7}$, though in this case $\beta$ must be modified
due to strong interactions between the leading quark and the medium,
which would distort the relation in Eq.(\ref{eqn:cos}).
The most important but least known factor is whether
the medium with $\delta n > 0$ could emerge or not
during the phase transition. However, whatever the mechanism is,
as long as the Cherenkov rings above $\sim$100 MeV associated with
preexisted charged leptons were observed,
one would be able to clearly state the dominant carriers of the medium are not pions
but light quarks apart from the complexity of the gluonic matter, since charged leptons
see only electromagnetic aspect of the QGP.

The estimated photon energy between Eq.(\ref{eqn:wp}) and (\ref{eqn:wppi})
is experimentally measurable even in the present detector design such as PHENIX
detector at RHIC \cite{PHENIXHI}\cite{PHENIXSpin} and ALICE detector at LHC\cite{ALICE},
both of which are capable of measuring photons as well as
charged leptons and hadrons.
Even with huge photon background conditions,
it would be feasible to reconstruct Cherenkov rings if we utilize special tools
such as wavelet analysis\cite{Dremin:Ring}, as long as the emission angle is
reasonably large enough compared to the granularity of photon detector segments.
If high $p_t$ light quarks are regarded as leading particles in a medium,
the Cherenkov rings can not be easily reconstructed. In this case one may
seek the unexpected fluctuations between high $p_t$ hadrons and photons
well beyond the fluctuations expected from jet fragmentation.
Unfortunately, in the case where $\delta n$ is quite small,
the photons are collinearly emitted with the penetrating charged
particles, then it would be difficult to show the existence of rings.
However, even in such a case, $E/p$, the fraction of measured energy $E$ in electromagnetic
calorimeters to measured momentum $p$ in tracking devices for each high $p_t$ charged
particle could be a good observable. 
If $E/p$ values of high $p_t$ charged particles are increased
due to collinearly emitted photon energy compared to fluctuations of
expected $E/p$ from accidentally associated photon energies
in the high multiplicity environment of heavy ion collisions,
it would be a hint of the existence of Cherenkov radiations.

In summary, the observation of Cherenkov photon radiation of the order
of 100 MeV associated with fast charged leptons from the
hot dense medium is a definite signature of Quark Plasma formation.
Although the observable focuses on only the electrical charge aspect,
it evidently indicates the lightness of carriers in the plasma.
It also opens a way to investigate the dispersion property
of a formed hot dense medium by measuring emission angles and
the wavelengths of Cherenkov photons. The cut-off or upper limit on the
energy of the emitted Cherenkov photon is within the measurable
energy range covered by the existing detector designs.
Although the observation of Cherenkov rings is the clearest signature,
photon and high $p_t$ hadron correlations or the large fraction
of measured energy to measured momentum of charged particles
indirectly indicates the existence of Cherenkov radiations.
I hope that more quantitative discussion on
$\delta n$ including the color charge aspect of the plasma
would follow this letter in connection with this observable.

\section*{Acknowledgements}
I would like to thank Tomoaki Nakamura and
Michael J. Tannenbaum for their discussions with me,
by which I happened to reach this idea.

\bibliographystyle{h-elsevier}
\bibliography{ref}
\end{document}